\documentclass[twocolumn,preprintnumbers,amsmath,aps]{revtex4}
\usepackage[dvips]{graphicx}
\usepackage{dcolumn}
\usepackage{bm}
\usepackage{color}
\usepackage{multirow}
\begin{document}
\def\eq#1{Eq.\hspace{1mm}(\ref{#1})}
\def\fig#1{Fig.\hspace{1mm}\ref{#1}}
\def\tab#1{Tab.\hspace{1mm}\ref{#1}}
\preprint{}
\title{
Nonadiabatic superconductivity in Li-intercalated hexagonal boron nitride bilayer}
\author{K. A. Szewczyk$^{\left(1\right)}$}
\email{kamila.szewczyk@ajd.czest.pl}
\author{I. A. Domagalska$^{\left(2\right)}$}
\author{A. P. Durajski$^{\left(3\right)}$}
\author{R. Szcz{\c{e}}{\'s}niak$^{\left(1,3\right)}$}
\affiliation{$^1$ Division of Theoretical Physics, Jan D{\l}ugosz University in Cz{\c{e}}stochowa, Ave. Armii Krajowej 13/15, 42-200 Cz{\c{e}}stochowa, Poland}
\affiliation{$^2$ Quantum Optics and Engineering Division, Faculty of Physics and Astronomy, University of Zielona G{\'o}ra, 
                  Prof. Z. Szafrana 4a, 65-516 Zielona G{\'o}ra, Poland}
\affiliation{$^3$ Division of Physics, Cz{\c{e}}stochowa University of Technology, Ave. Armii Krajowej 19, 42-200 Cz{\c{e}}stochowa, Poland}
\date{\today} 
\begin{abstract}
In the case of Li-intercalated hexagonal boron nitride bilayer (Li-hBN), the vertex corrections of electron-phonon interaction cannot be omitted. 
This is evidenced by the very high value of the ratio $\lambda\omega_{D}/\varepsilon_{F}\sim 0.46$, where $\lambda$ is the electron-phonon coupling constant, $\omega_{D}$ is the Debye frequency, and the symbol $\varepsilon_{F}$ represents the Fermi energy. 
Due to the nonadiabatic effects, the phonon-induced superconducting state in Li-hBN is characterized by the much lower value of critical temperature (\mbox{$T^{\rm LOVC}_{C}\in\{ 19.1, 15.5, 11.8\}$~K}, for \mbox{$\mu^{\star}\in \{0.1, 0.14, 0.2\}$}), than would result from calculations not taking this effect into account: \mbox{$T^{\rm ME}_{C}\in\{ 31.9, 26.9, 21\}$~K}. From the technological point of view, the low value of $T_{C}$ limits the possible applications of Li-hBN superconducting properties. The calculations were carried out under the classic Migdal-Eliashberg formalism (ME) and the Eliashberg theory with the lowest-order vertex corrections (LOVC).

\vspace*{0.4cm}
\end{abstract}
\maketitle

\vspace{0.5cm} 

\section{Introduction}
Low-dimensional systems: graphene \cite{Ohta2006A, Profeta2012A, Pesic2014A, Guzman2014A, Margine2016A}, silicene \cite{Wan2013A}, borophene \cite{Gao2017A, Liao2017A}, and phosphorene \cite{Liu2014A, Ge2015A, Shao2014A} are mechanically stable only when placed on the substrate \cite{Kolesov2019A, Wang2017A, Slawinska2010A}. The substrate should be selected so that it changes the physical properties of the low-dimensional system as little as possible. In the case of graphene, the following were used as the substrate material: Co \cite{Hamilton1980A}, Ni \cite{Shikin2000A, Rosei1983A, Shikin1998A, Dedkov2001A}, Ru \cite{Brugger2009A, Moritz2010A}, Pt \cite{Land1992A, Starr2006A}, SiC \cite{Forbeaux1998A, Mendes2012A, Hass2008A}, and $\rm SiO_{2}$ \cite{Chen2008A, Lee2008A, Geringer2009A}. Unfortunately, the obtained experimental data showed that the incompatible crystalline structure of the above materials leads to significant suppression of the carrier mobility of graphene \cite{Wang2017A, Ponomarenko2011A}.

It is now assumed that the best substrate for graphene is the hBN system with the honeycomb crystal structure (boron (B) and nitrogen (N) atoms alternating between hexagonal lattice nodes). In crystalline form, hBN was synthesized by Nagashima {\it et al.} in 1995 \cite{Nagashima1995A}. A decade later, the two-dimensional form of hBN was obtained at the University of Manchester \cite{Novoselov2005A}.

Graphene and hBN monolayer have the extremely similar crystal lattice structure. Their compatibility is estimated at $98.5$~\% \cite{Starr2006A}. 
In the case of the graphene/hBN composite, the homogeneous distribution of charge on the graphene surface is observed. Note that this result is radically different from the data obtained for graphene/$\rm SiO_{2}$ \cite{Decker2011A}. In addition, hBN monolayer exhibits the high temperature stability. It is characterized by the low dielectric constant ($\varepsilon\sim 3-4 $), and the high thermal conductivity \cite{Giovannetti2007A}. The band gap of hBN is about $5.9$~eV \cite{Dean2010A}. Furthermore, which is also important the hBN is non-toxic.

It is worth noting that graphene on hBN substrate was used to make transistors device with high mobility \cite{Dean2010A}, with the help of which the quantum Hall effect was observed. The heterojunction with two graphene layers \cite{Ponomarenko2011A}, and superlattice structures \cite{Britnell2012A, Haigh2012A, Dean2012A} were also constructed. The graphene/hBN heterojunction devices allowed detection of the Hofstadter Butterfly phenomenon \cite{Ponomarenko2013A, Dean2013A}.

The hBN structure in both layered and volume forms has the very wide energy gap, which makes it the insulator \cite{Wang2017A, Shimada2017A}. Therefore, for the long time this material was not associated with superconductivity phenomenon. The situation changed when it was suggested 
that hBN intercalation of the lithium induces the transition to metallic state \cite{Altintas2011A}. Let us note that quasi-two-dimensional superconducting systems are currently being intensively studied for their possible applications in the nano-superconducting quantum interference 
devices \cite{Fatemi2018A} and the quantum information technology \cite{Fagaly2006A, Komatsu2018A}. 

Currently, the most promising research seems to be the properties of the superconducting state in the Li-intercalated hexagonal boron nitride bilayer (Li-hBN) compound. Based on DFT calculations, it has been shown that the critical temperature ($T_ {C}$) of the superconductor-metal phase transition is about $25$~K \cite{Shimada2017A}, for the Coulomb pseudopotential $\mu^ {\star}=0.14$ (identical to the experimental value $\mu^{\star}$ obtained for graphene 
\cite{Ludbrook2015A}).
The expected value of $T_{C}$ proved to be much higher than the maximum temperature that was achieved in graphene intercalated with alkali metals ($T_{C}=8.1$~K in Ca-intercalated bilayer graphene) \cite{Margine2016A}. As well as in other superconducting low-dimensional structures: $T_ {C}\sim 20$~K for Li- and Na-intercalated blue phosphorene bilayer \cite{Zhang2016A}, $T_{C}\sim 16.5 $~K for Li-intercalated black phosphorene bilayer \cite{Huang2015A}, and $T_ {C}\sim 10$~K for Li-$ {\rm MoS_{2}}$ bilayer \cite{Huang2016A}, {\it etc}. 

The obtained result for Li-hBN is explained by the relatively high value of the electronic density of states at the Fermi level and the significant contribution to the pairing interaction from the inter-layered electron-phonon coupling \cite{Shimada2017A}. This is due to the formation of characteristic bonds connecting two boron atoms in the upper and lower layers of hBN, which results from the poor electronegativity of boron atoms.

From the experimental side, we have the results of research conducted in 2019 by S. Moriyama {\it et al.} \cite{Moriyama2019A}. The existence of the superconducting state has been observed in the system consisting of the non-twisted bilayer graphene (BLG) and the hexagonal boron nitride layers (hBN/BLG/hBN). The following characteristic temperatures were obtained: $T^{\rm onset}\sim 50$ ~K, $T^{\star}\sim 30$~ K, and $T_{\rm BKT}=14$~K, which correspond the onset of superconductivity ($90$\% of the total transition/normal resistance), the crossover to superconductivity ($50$\% of the normal resistance) and the confinement of vortices, respectively.

The important question is whether the Li-hBN bilayer system can get as high critical temperature as suggest the DFT calculations ($T_{C}=25$~K) \cite{Shimada2017A}. In our opinion not, due to the fact that the electron-phonon interaction in Li-hBN must be taken into account together with vertex corrections. This is demonstrated by the very high ratio: $\lambda\omega_{D}/\varepsilon_{F}\sim 0.46$, where $\lambda=1.17$ is the electron-phonon coupling constant, $\omega_ {D}=165.56$~meV is the Debye frequency, and the symbol $\varepsilon_ {F} =417.58$~meV represents the Fermi energy \cite{Shimada2017A}.

For this reason, in the presented paper, we characterized the properties of the superconducting state in Li-hBN bilayer in the framework of Eliahberg formalism, which includes the vertex corrections of electron-phonon interaction \cite{Freericks1997A}. We compared the results with those obtained using the standard Migdal-Eliashberg theory \cite{Eliashberg1960A}. Note that the use of Eliashberg formalism is associated with the high value of the electron-phonon coupling constant $\lambda$, which characterizes the superconducting state in Li-hBN \cite{Shimada2017A}. Let us remind that the BCS theory gives the correct results only in the weak-coupling limit, where $\lambda<0.3$ \cite{Bardeen1957A, Bardeen1957B}.

\section{Theoretical model}

The  classical {\bf M}igdal-{\bf E}liashberg (ME) formalism \cite{Migdal1958A, Eliashberg1960A} represents the natural generalization of BCS theory (the first microscopic theory of superconducting state) \cite{Bardeen1957A, Bardeen1957B}. This generalization consists in taking into account the retardation and strong-coupling effects of the electron-phonon interaction, which is responsible for the condensation of electrons in the Cooper pairs \cite{Cooper1956A}. As part of Eliashberg formalism, the electron-phonon interaction is quantified by the so-called Eliashberg function ($\alpha^{2}F\left(\omega\right)$). The form of Eliashberg function for the specific physical system can be determined theoretically by the DFT method \cite{Giustino2017A}, or experimentally using the data provided by the tunnel experiment \cite{McMillan1965A, Yanson1974A}. The electron correlations (the screened Coulomb interaction) are modelled parametrically defining the so-called Coulomb pseudopotential ($\mu^{\star}$) \cite{Morel1962A}. The function $\alpha^{2}F\left(\omega\right)$ and $\mu^{\star} $ are the only input quantities of Eliashberg equations characterizing the properties of the system in which induction of the superconducting state is expected.

The classical Eliashberg equations are thoroughly discussed in the literature \cite{Carbotte1990A}. They allow the self-consistent determination of the superconducting order parameter ($\Delta_{n}=\Delta\left(i\omega_ {n}\right)$) and the wave function renormalization factor ($Z_{n}=Z\left(i\omega_ {n}\right) $), with the accuracy of the second order relative to the electron-phonon coupling function ($g$). The symbol $\omega_ {n}=\pi k_ {B}T\left (2n +1\right)$ defines the fermionic Matsubara frequency. In the case of the phonon-induced superconducting state, the limitation of considerations to the order of $g^{2}$ is justified by the Migdal theorem \cite{Migdal1958A}. The Migdal theorem applies when the ratio $\lambda\omega_ {D}/\varepsilon_ {F}$ is in the order of $0.01$. This means that the energy of the phonons is so small that the vertex corrections for the electron-phonon interaction are irrelevant.

 Based on DFT calculations, the value of the $\lambda\omega_{D}/\varepsilon_{F} $ ratio for Li-hBN is equal to $0.46$. For this reason, the superconducting state in Li-hBN cannot be quantified in the classical Eliashberg theory. Note that the unusually high value of the $\lambda\omega_ {D}/\varepsilon_ {F} $ ratio for Li-hBN is related to the fact that the physical system is quasi-two-dimensional. In the case of the bulk superconductor, the width of the electron band is significantly broadened, which results in the increase in the Fermi energy ($\varepsilon_{F}=1.63$~eV). In addition, the electron-phonon coupling constant ($\lambda=0.66$) decreases. As the result, $\lambda\omega_ {D}/\varepsilon_ {F}$ is just $0.07$. The calculations carried out by us within the Migdal-Eliashberg formalism prove that the superconducting state in the bulk system have the significantly lower critical temperature value than in the quasi-two-dimensional system. In particular, we received: $T_{C}\in\{14.01, 8.64, 4.6 \}$~K, for $\mu^{\star}\in\{0.1, 0.2, 0.3 \} $.

To realize how high the value of $\lambda\omega_ {D}/\varepsilon_ {F} $ for Li-hBN is, it is enough to note that for the  Li-${\rm MoS_{2}} $ bilayer, we get $\lambda\omega_ {D}/\varepsilon_ {F} =0.15$  \cite{Huang2016A}. In the bilayer of black and blue phosphorus intercalated with lithium, $\lambda\omega_{D}/\varepsilon_{F}$ is $0.05$ and $0.1$, respectively \cite{Huang2015A, Zhang2016A}. It is worth noting that the value of the parameter $\lambda\omega_ {D}/\varepsilon_ {F}$ at the level of $0.09$ causes the noticeable modification of the superconducting state properties, as in the case of ${\rm LiC_ {6 }} $, where $T_{C}\sim 6$~K \cite {Profeta2012A, Ludbrook2015A, Zheng2016A, Szczesniak2019A}. 

Therefore, to describe the superconducting phase in Li-hBN, we used the Eliashberg equations derived with the accuracy of the fourth order relative to $g$ ({\bf L}owest-{\bf O}rder {\bf V}ertex {\bf C}orrections (LOVC)). These equations were derived in \cite{Freericks1997A} by Freericks {\it et al.}, where they were used to analyze the properties of the superconducting state inducing in lead. They take the form ($A=1$):
\begin{widetext}
\begin{eqnarray}
\label{r01-II}
\varphi_{n}&=&\pi k_{B}T\sum_{m=-M}^{M}
\frac{\lambda_{n,m}-\mu_{m}^{\star}}
{\sqrt{\omega_m^2Z^{2}_{m}+\varphi^{2}_{m}}}\varphi_{m}\\ \nonumber
&-&
A\frac{\pi^{3}\left(k_{B}T\right)^{2}}{4\varepsilon_{F}}\sum_{m=-M}^{M}\sum_{m'=-M}^{M}
\frac{\lambda_{n,m}\lambda_{n,m'}}
{\sqrt{\left(\omega_m^2Z^{2}_{m}+\varphi^{2}_{m}\right)
       \left(\omega_{m'}^2Z^{2}_{m'}+\varphi^{2}_{m'}\right)
       \left(\omega_{-n+m+m'}^2Z^{2}_{-n+m+m'}+\varphi^{2}_{-n+m+m'}\right)}}\\ \nonumber
&\times&
\left[
\varphi_{m}\varphi_{m'}\varphi_{-n+m+m'}+2\varphi_{m}\omega_{m'}Z_{m'}\omega_{-n+m+m'}Z_{-n+m+m'}-\omega_{m}Z_{m}\omega_{m'}Z_{m'}
\varphi_{-n+m+m'}
\right],
\end{eqnarray}
and
\begin{eqnarray}
\label{r02-II}
Z_{n}&=&1+\frac{\pi k_{B}T}{\omega_{n}}\sum_{m=-M}^{M}
\frac{\lambda_{n,m}}{\sqrt{\omega_m^2Z^{2}_{m}+\varphi^{2}_{m}}}\omega_{m}Z_{m}\\ \nonumber
&-&
A\frac{\pi^{3}\left(k_{B}T\right)^{2}}{4\varepsilon_{F}\omega_{n}}\sum_{m=-M}^{M}\sum_{m'=-M}^{M}
\frac{\lambda_{n,m}\lambda_{n,m'}}
{\sqrt{\left(\omega_m^2Z^{2}_{m}+\varphi^{2}_{m}\right)
       \left(\omega_{m'}^2Z^{2}_{m'}+\varphi^{2}_{m'}\right)
       \left(\omega_{-n+m+m'}^2Z^{2}_{-n+m+m'}+\varphi^{2}_{-n+m+m'}\right)}}\\ \nonumber
&\times&
\left[
\omega_{m}Z_{m}\omega_{m'}Z_{m'}\omega_{-n+m+m'}Z_{-n+m+m'}+2\omega_{m}Z_{m}\varphi_{m'}\varphi_{-n+m+m'}-\varphi_{m}\varphi_{m'}\omega_{-n+m+m'}Z_{-n+m+m'}
\right],
\end{eqnarray}
\end{widetext}
where for $A=0$, we get the classic Migdal-Eliashberg equations. The order parameter is given by the formula: $\Delta_{n} =\varphi_{n}/Z_{n}$. The symbol $\lambda_{n, m}$ means the pairing kernel for the electron-phonon interactions:
\begin{equation}
\label{r03-II}
\lambda_{n,m}=2\int_0^{\omega_{D}}d\omega\frac{\omega}{\omega ^2+4\pi^{2}\left(k_{B}T\right)^{2}\left(n-m\right)^{2}}\alpha^{2}F\left(\omega\right).
\end{equation}
The Coulomb pseudopotential function is: $\mu_{n}^{\star}=\mu^{\star}\theta \left(\omega_{c}-|\omega_{n}|\right)$, where $\theta\left(x\right)$ is the Heaviside function, and $\omega_{c}$ represents the cut-off frequency ($\omega_{c}=3\omega_{D}=496.7$~meV).  

Freericks equations allow to determine the values of the order parameter and the wave function renormalization factor in the self-consistent manner, which is undoubtedly their great advantage. These are isotropic equations, which means that the self-consistent procedure does not apply to the electron momentum ($ {\bf k} $). From the physical point of view, however, this should not be significant, because the phonon-induced superconducting state is highly isotropic \cite{Carbotte1990A}. The situation would of course change radically if, in addition, the strong electron correlations had to be taken into account. Nevertheless, the Eliashberg equations including vertex corrections and explicit dependence on $ {\bf k}$ are also given in the literature \cite{Grimaldi1995A, Pietronero1995A, Grimaldi1995B}. These equations were derived in the context of research on the superconducting state in the fullerene systems \cite{Pietronero1992A, Pickett1993A}, in the high-$T_{C}$ cuprates \cite{Uemura1991A, Uemura1992A, Ambrumenil1991A}, in the heavy fermion compounds \cite{Wojciechowski1996A}, and in the superconductors under high magnetic fields \cite{Goto1996A}. Unfortunately, due to enormous mathematical difficulties, their full self-consistent solutions are still unknown ($\Delta_{n,\bf{k}}$ and $Z_{n,\bf{k}}$).

It is also worth noting that Freericks equations have recently been successfully used to analyze the superconducting state with high critical temperature values in compounds such as ${\rm PH_{3}}$ ($T_{C}\sim 80$~K), ${\rm H_{3}S}$ ($T_ {C}\sim 200$~K) \cite {Durajski2016A} 
and ${\rm H_{2}S}$ ($T_{C}\sim 35$~K) \cite{Kostrzewa2018A}. 

From the mathematical point of view, the Eliashberg equations are solved in the self-consistent manner, taking into account the correspondingly large number of fermionic Matsubara frequencies \cite{Szczesniak2006A, Wiendlocha2016A}. In our considerations, we assumed that the number $M$ is $4000$, which ensured the appropriate convergence of solutions of Eliashberg equations for the temperature higher or equal to $T_{0}=4$~K. Due to the lack of experimental data, in the examined physical system, we took into account the Coulomb pseudopotential in the range from $0.1$ to $0.2$, where the value of $0.14$ was already considered in the paper \cite{Shimada2017A}.

\section{Results}
%
\begin{figure}
\includegraphics[width=\columnwidth]{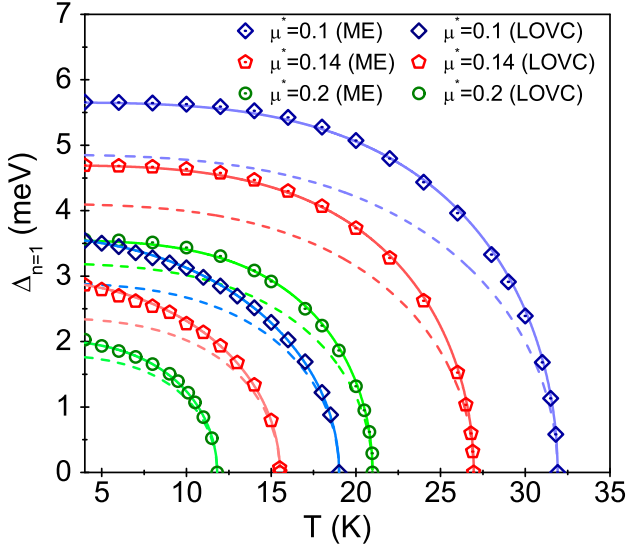}
\caption{The order parameter as a function of temperature. ME model - symbols with the dot, LOVC model - the empty symbols. Adopted, $\mu^{\star}\in\{0.1, 0.14, 0.2 \} $. The solid lines represent the parameterization of numerical results using \eq {r03-III}. The dashed lines were obtained as part of the BCS theory (the mean-field theory).}
\label{f01}  
\end{figure}
\begin{figure}
\includegraphics[width=0.9\columnwidth]{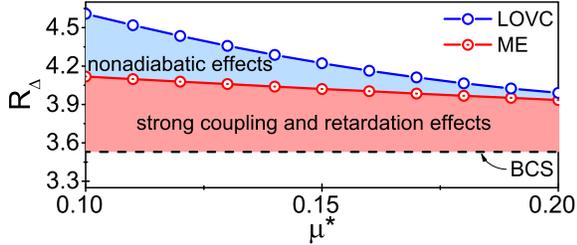}
\caption{The values of the ratio $R_{\Delta}$ as a function of the Coulomb pseudopotential. The results obtained under the model: LOVC, ME, and BCS.}
\label{f02}  
\end{figure}

In \fig{f01}, we plotted the dependence of the order parameter on temperature. Note that under the imaginary axis formalism, it is assumed that the physical value of the order parameter is $\Delta_{n=1}$. In the classic ME model, we obtained the following critical temperature values: $T^{\rm ME}_{C}\in\{31.9, 26.9, 21\}$~K, respectively for $\mu^{\star}\in\{0.1, 0.14, 0.2\} $. Comparing the obtained results with the results taking into account the impact of the vertex corrections ($T^{\rm LOVC}_{C}\in\{19.1,15.5,11.8\}$~K), we find that the nonadiabatic superconducting state in Li-hBN have the much lower value of $T_{C}$ than it would follow from the ME model. 

The observed effect of lowering the critical temperature value does not only result from the static corrections ($Stat.$), whose good measure is the ratio $m=\omega_{D}/\varepsilon_ {F}=0.4$ (Migdal parameter). It should also be associated with dynamic corrections modeled by the explicit dependence of the order parameter and the wave function renormalization factor on the Matsubara frequency.

Based on the results of \cite{Grimaldi1995B, Combescot1990A}, the impact of static vertex corrections on critical temperature can be estimated using the formula:
\begin{equation}
\label{r01-III} 
T^{Stat.}_{C}=\chi T^{AD}_{C},
\end{equation}
where the symbol $T^{AD}_{C}$ means the critical temperature value calculated on the basis of the Allen-Dynes formula \cite{Allen1975A}. The input from the static part of the vertex corrections has the form:
\begin{equation}
\label{r02-III} 
\chi=\frac{1}{m+1}e^{\frac{2m-1}{2\left(m+1\right)}}.
\end{equation}

The good measure of the impact of the dynamic part of the vertex corrections on the critical temperature value is: $D=\left[(T^{Stat.}_{C}-T^{LOVC}_{C})/(T^{ME}_{C}-T^{LOVC}_{C})\right]\cdot 100$\%.

We collected the results in \tab{t01}. As one can see, the static part of the vertex corrections is responsible for $80$-$90$~\% of the difference in the $T_ {C} $ predicted by the ME and LOVC models.   

\begin{table*}
\caption{\label{t01} The critical temperature estimated in the LOVC model, in the ME model, using the Allen-Dynes formula \cite{Allen1975A}, and in the analytical model including static corrections ($T^{Stat.} _ {C} $). Additionally, the values of the $D$ parameter were given.}
\begin{ruledtabular}
\begin{tabular}{|c||c|c|c|c|c|}
               &                    &                  &                  &                     &                                              \\
 $\mu^{\star}$ & $T^{LOVC}_{C}$~(K) & $T^{ME}_{C}$~(K) & $T^{AD}_{C}$~(K) & $T^{Stat.}_{C}$~(K) &  $D$\%                                       \\
               &                    &                  &                  &                     &                                              \\
\hline
               &                    &                  &                  &                     &                                              \\
     0.1       & {\bf 19.1}         &  31.9            &  32.2            &   {\bf 21.4}        &  {\bf 18}                                    \\
               &                    &                  &                  &                     &                                              \\
     0.14      & {\bf 15.5}         &  26.9            &  26.7            &   {\bf 17.8}        &  {\bf 20.2}                                  \\
               &                    &                  &                  &                     &                                              \\
     0.2       & {\bf 11.8}         &  21              &  19.4            &   {\bf 12.9}        &  {\bf 12}                                    \\
               &                    &                  &                  &                     &                                              \\
\end{tabular}
\end{ruledtabular}
\end{table*}

The numerical results obtained from the Eliashberg equations can be parameterized using the formula \cite{Eschrig2001A}:

\begin{equation}
\label{r03-III} 
\Delta(T)=\Delta(0)\sqrt{1-(T/T_{C})^{\Gamma}},
\end{equation}
where $\Delta\left(0\right)=\Delta\left(T_ {0}\right)$. In the case of the LOVC model, we received $\Gamma\in\{2.17, 2.2, 2.8\}$, respectively for $\mu^{\star}\in\{0.1, 0.14, 0.2 \}$. The exponent $\Gamma$ for the classic ME approach differs significantly in values: $\Gamma\in\{3.45, 3.4, 3.45 \}$. 
The accuracy of analytical parameterization of the numerical results is presented in \fig{f01} (solid lines). In addition, the results obtained under the mean-field BCS model were marked using dashed lines. In this case, $\Delta (0)=1.76\cdot k_ {B}T_ {C}$ was adopted \cite{Bardeen1957A, Bardeen1957B}. 
The value of the exponent $\Gamma$ for the BCS model is equal to $3$ \cite{Eschrig2001A}.  

Note the differences in the shape of the curves corresponding to the parameterization of the Eliashberg results and the BCS theory. In the case of the ME model, the differences result only from the retardation and strong-coupling effects correctly taken into account in the ME formalism. 
These effects can be characterized by calculating the value of the ratio $r=k_ {B} T_{C}/\omega_ {\ln} $, where the symbol $\omega_ {\ln}=\exp\left[\frac{2}{\lambda}\int^{+\infty}_{0} d\Omega\frac{\alpha^{2}F\left(\Omega\right)}{\Omega}\ln\left(\Omega\right)\right]=28.98$~meV is called 
the logarithmic phonon frequency \cite{Allen1975A}. The $r$ parameter for Li-hBN is $r^{\rm ME}\in\{0.095, 0.08, 0.062 \} $ or 
$r^{\rm LOVC}\in\{0.057, 0.046, 0.035 \}$, respectively for $\mu^{\star}\in\{0.1, 0.14, 0.2 \} $. This means that the effects considered are significant even when we consider the vertex corrections for the electron-phonon interaction. Also note that the retardation and strong-coupling effects for Li-hBN are of the same order as in Li-${\rm MoS_{2}}$ bilayer compounds \cite{Huang2016A}, Li-black phosphorene bilayer \cite{Huang2015A}, and Li-blue phosphorene bilayer \cite{Zhang2016A}: $0.068$, $0.094$, and $0.099$ (these results were obtained for $T_{C}$ determined from Allan-Dynes formula \cite{Allen1975A} assuming $\mu^{\star}=0.1$). In the BCS limit, the Eliashberg equations predict $r\rightarrow 0$.

In the LOVC theory, we take into account the vertex corrections as well as retardation and strong-coupling effects, as a result the differences between the Eliashberg parameterization curves and the BCS curves noticeably increase. The good measure of this effect is the value of the ratio 
$R_{\Delta}=2\Delta(0)/k_ {B}T_ {C}$. For the Li-hBN system, we received: 
$R^{\rm LOVC}_ {\Delta}\in\{4.6, 4.29, 3.99\} $ and $R^{\rm ME}_{\Delta}\in\{4.12, 4.04, 3.9 \} $. 
It should be emphasized that in the case of BCS theory, the value of $R_ {\Delta}$ is $3.53$ - it is the universal constant of the model \cite{Bardeen1957A, Bardeen1957B} . The results obtained for 
$\mu^{\star}\in\left<0.1,0.2\right>$ are presented in \fig{f02}. One cannotice the interesting effect. 
Namely, with the increase of the depairing electron correlations, the impact of vertex corrections on the $R_ {\Delta}$ ratio value decreases, so for $\mu^{\star}\sim 0.2$ the parameter 
$R^{\rm LOVC}_{\Delta}$ differs only slightly from $R^{\rm ME}_ {\Delta} $.

\begin{figure}
\includegraphics[width=\columnwidth]{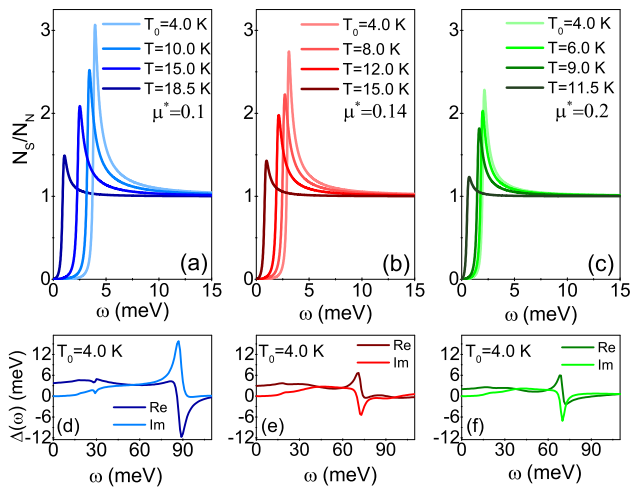}
\caption{(a) - (c) The normalized density of states for selected temperature.
         (d) - (f) The form of the order parameter on the real axis calculated for $T=4$~K. 
         The results were obtained in the framework of LOVC model.         }
\label{f03}  
\end{figure}

Having the full dependence of the order parameter on the Matsubara frequency, we determined the normalized density of states: 
\begin{equation}
\label{r04-III}
\frac{N_{S}\left(\omega\right)}{N_{N}\left(\omega\right)}={\rm Re}\left[\frac{|\omega-i\delta|}
{\sqrt{\left(\omega-i\delta\right)^{2}-\left(\Delta\left(\omega\right)\right)^{2}}}\right], 
\end{equation} 
where the pair breaking parameter $\delta$ equals $0.15$~meV. We calculated the value of $\Delta\left(\omega\right)$ by continuing the functions $\Delta_ {n}$ on the real axis \cite{Beach2000A}. 
The results obtained under the LOVC approach for $N_{S}\left(\omega\right/N_{N}\left(\omega\right)$ are collected in \fig{f03} (a)-(c). The presented curves can also be determined on the basis of the data obtained using the tunneling junction. Hence, any experimental results directly relate to the predictions 
of Eliashberg formalism taking into account the effect of vertex corrections. Additionally, in \fig{f03} (d)-(f) we plotted the form of the order parameter on the real axis ($T=4$~K). The real part of the function $\Delta\left(\omega\right)$ specifies the physical value of the order parameter, which can be calculated using the equation \cite{Carbotte1990A}: 
$\Delta\left(T\right)={\rm Re}\left [\Delta\left(\omega =\Delta\left(T\right)\right)\right] $. In the present case, we obtained values that differ from $\Delta_{n=1}$ not more than $10^{-2}$\%. This result proves that the analytical continuation was correct. On the other hand, the imaginary part of the $\Delta\left(\omega\right)$ function determines the damping effects. One can see that at low frequencies, where
${\rm Im}\left[\Delta\left(\omega\right)\right]=0$, these effects do not occur. From the physical point of view, this means the infinite lifetime of the Cooper pairs. Above the frequency 
$\omega\sim 15$~meV, both the real and imaginary part of the order parameter function have the complicated course. This fact results directly from the complicated shape of the Eliashberg function, which models the electron-phonon interaction in the Li-hBN system.  
\begin{figure}
\includegraphics[width=\columnwidth]{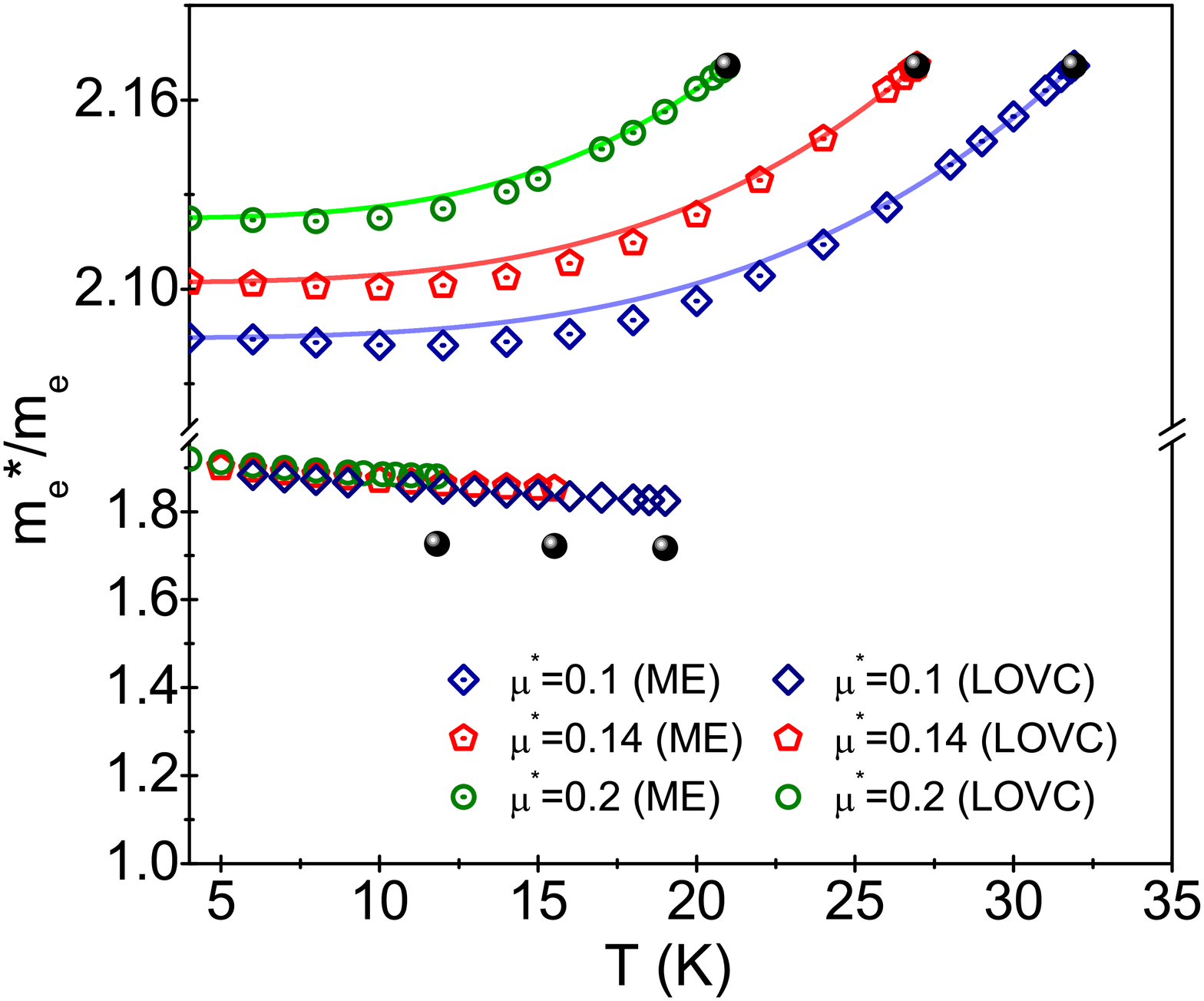}
\caption{The ratio of the electron effective mass to the electron band mass as a function of temperature. 
         The results were obtained in the framework of LOVC model, ME model, and \eq{r06-III}. 
         The lines for ME results can be reproduced using the formula:
          $m_{e}^{\star}\slash m_{e}=\left[Z_{n=1}\left(T_{C}\right)-Z_{n=1}\left(0\right)\right]\left(T/T_{C}\right)^{\Gamma}
           +Z_{n=1}\left(0\right)$, where $Z_{n=1}\left(0\right)=Z_{n=1}\left(T_{0}\right)$.}
\label{f04}  
\end{figure}

Let's discuss the effect of vertex corrections on the electron band mass ($m_{e}$). 
To do this, it is necessary use the formula: $m^{\star}_{e}/m_{e}=Z_{n=1} $, where the symbol $m^{\star}_{e}$ represents the effective electron mass.

The results obtained on the basis of Eliashberg equations are presented in \fig{f04}. It is easy to see that the effective mass of the electron is almost twice as high as the electron band mass, with $m^{\star}_{e}$ very slightly dependent on the temperature. The vertex corrections lower the value of $m^{\star}_{e}$ compared to the value predicted under ME formalism. If the temperature equals the critical temperature, this effect can be characterized analytically. To do this consider the \eq{r02-II}, which for $Z_{n=1} $ takes the form:
\begin{eqnarray}
\label{r05-III}
& &Z_{n=1}=1+\lambda\sum^{M}_{m=-M}{\rm sgn}\left(\omega_{m}\right)\\ \nonumber
&-&\lambda^{2}\frac{\pi^{2}}{4}\frac{k_{B}T_{C}}{\varepsilon_{F}}\sum^{M}_{m,m'=-M}
{\rm sgn}\left(\omega_{m}\right){\rm sgn}\left(\omega_{m'}\right){\rm sgn}\left(\omega_{m+m'}\right).
\end{eqnarray}
Hence: 
\begin{equation}
\label{r06-III}
Z^{LOVC}_{n=1}-Z^{ME}_{n=1}=-\lambda^{2}\left(\frac{\pi}{4}\frac{\omega_{D}}{\varepsilon_{F}}+\frac{\pi^{2}}{2}\frac{k_{B}T_{C}}{\varepsilon_{F}}\right),
\end{equation}
where $Z^{ME}_{n=1}=1+\lambda$. Based on \eq{r06-III}, it can be concluded that the lowest-order vertex corrections lower the effective mass value of electron the stronger the $\lambda$ and 
$\omega_{D}$ are higher, but it should be noted that in this case, the critical temperature also increases. 
The values of $Z^{LOVC}_{n=1}$ and $Z^{ME}_{n=1}$, calculated on the basis of \eq{r06-III}, have been marked on \fig{f04} using black spheres. We received the good agreement between numerical and analytical results.

\section{Summary and discussion of results}

To sum up, the superconducting state in Li-hBN is induced by the electron-phonon interaction, which is characterized by the rare, very high value of the ratio $\lambda\omega_ {D}/\varepsilon_{F} = 0.46$. This means that the thermodynamic properties of the superconducting phase should be determined using formalism explicitly including the vertex corrections. Note that the very high value of the $\lambda\omega_{D}/\varepsilon_ {F}$ ratio is related to the quasi-two-dimensionality of the system under consideration \cite{Shimada2017A}.

In the paper, we showed that the nonadiabatic effects significantly lower the critical temperature ($T^{\rm LOVC}_{C}\in\{19.1, 15.5, 11.8\}$~K), compared to the results obtained in the framework of the Migdal-Eliashberg theory: $T^{\rm ME}_{C}\in\{31.9, 26.9, 21\}$~K, for $\mu^{\star}\in\{0.1, 0.14, 0.2 \}$. In our opinion, there is no reason to believe that the critical temperature in Li-hBN exceeds $20$~K, which certainly limits applications of the tested material.

Note that the low values of $T_{C}$ occur in principle in the whole family of systems, where the honeycomb crystal structure plays the important role \cite{Margine2016A, Zhang2016A, Huang2015A, Huang2016A}. This structure, however fundamental for the properties of graphene, is unfavorable for the superconducting state. The reason for this is that van Hove singularity in the electronic density of states is considerably distant form the Fermi level \cite{VanHove1953A}. This is not the case for the square lattice, where the van Hove singularity is very close or even at the Fermi level, which means that the value of $T_{C}$ can increase by the order of magnitude \cite{Szczesniak2012A}.
\begin{figure}
\includegraphics[width=\columnwidth]{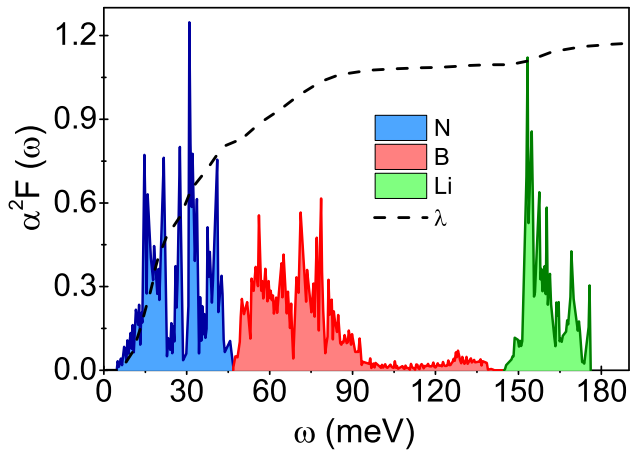}
\caption{ The Eliashberg function $\alpha^{2}F\left(\omega\right)$ and electron-phonon coupling function 
$\lambda\left (\omega'\right)=2\int^{\omega'}_{0} d\omega\alpha^{2}F\left(\omega\right)/\omega$ for Li-hBN. The results were obtained in the paper \cite{Shimada2017A}. The figure also indicates the contributions from nitrogen, boron and lithium: $\lambda^{\rm N}=0.82$, $\lambda^{\rm B}=0.25$ and 
$\lambda^{\rm Li}=0.1$, where $\lambda^{\rm N}+\lambda^{\rm B} + \lambda^ {\rm Li}=1.17$.} 
\label{f05}  
\end{figure}

However, the natural question arises whether the results obtained for Li-hBN suggest the alternative way to obtain material with the higher value of $T_{C}$. We believe that there are potentially such possibilities. To do this, consider the form of the Eliashberg function for Li-hBN (see \fig{f05}). It can easily be seen that the Eliashberg function consists of two clearly separated parts (the similar situation occurs in the case of hydrogen compounds \cite{Li2014A, Duan2014A}). In the low frequency range 
($\omega\in\left(4.59, 93.29 \right)$~meV) nitrogen and boron contributions are important. 
In the frequency range from $145.16$~meV to $176.13$~meV, the electron-phonon interaction associated with lithium atoms dominates. These frequency ranges are separable, with the Eliashberg function taking very small values in the range from $93.29$~meV to $145.16$~meV. The above facts suggest that the composition of the compound in question could be changed in such a way as to significantly increase the Eliashberg function values in the range from $93.29$~meV to $145.16$~meV. Most likely by appropriate doping of the starting compound. However, this is not the simple issue and requires DFT calculations.

Also striking is the possibility of substitution (at least partially) of lithium by hydrogen or boron and nitrogen by heavier elements. In the first case, the increase in critical temperature could be associated with the increase in Debye frequency ($T_{C}\sim\omega_ {D}$ - lower mass of the hydrogen nucleus in relation to the mass of the lithium nucleus: $\omega_{D}\sim 1/\sqrt{M}$). 
In the second case, the increase in $T_{C}$ could result from the increase in the electron-phonon coupling constant ($T_{C}\sim\exp\left(-1/\lambda\right)$ - contributions from heavy elements in the Eliashberg function located are in the low frequency range, which are potentially more significant for $\lambda$. 
To find out, just pay attention to the definition of the electron-phonon coupling constant: 
$\lambda=2\int^{\omega_{D}}_{0}d\omega\alpha^{2}F\left(\omega\right)/\omega$.

In the last paragraph, let us note that from the point of view of fundamental research on the phonon-induced superconducting state, the Li-hBN system seems to be very interesting because of 
the unusually high value of the ratio $\lambda\omega_ {D}/\varepsilon_ {F}$ - comparable to the value obtained for the fullerene compounds \cite{Pietronero1992A, Pietronero1995A}. Therefore, Li-hBN can be used to test the predictions of future theory that includes vertex corrections in the fully self-consistent manner (both Matsubara frequencies and the electron wave vector {\bf k}). We are currently investigating this issue extensively. Preliminary results for ME formalism can be found in the papers 
\cite{Szewczyk2018A, Szewczyk2019A}.

\acknowledgements{The authors would like to thank Nao H. Shimada, Emi Minamitani, and Satoshi Watanabe (University of Tokyo) for providing data on the Eliashberg function for Li-hBN bilayer, presented in \cite{Shimada2017A}, and for providing information on the electronic structure for bulk Li-hBN.}
\bibliography{bibliography}
\end{document}